\documentstyle[12pt]{article}
\begin{document}

\baselineskip=24pt plus 2pt
\hfill{NCKU-HEP/94-01}

\hfill{27 October 1994}
\begin{center}
\vspace{5mm}
{\large \bf
Spin Structure of the Proton\\}
\vspace{15mm}
Chung-Yi Wu
\vspace{5mm}

Department of Physics \\
National Cheng Kung University \\
Tainan, Taiwan 701, Republic of China \\
\end{center}
\vspace{15mm}
\begin{center}
{\bf ABSTRACT}
\end{center}
By assuming that there is no significant intrinsic polarization of the
gluon, we have computed the polarized quark contributions to the proton's
spin under SU(3) flavor symmetry breaking for the polarized sea and have
performed a global leading-order QCD fit to obtain the spin-dependent
quark distributions, which could be used as input for analyzing
lepton-hadron and hadron-hadron collisions.

\newpage
The measurement of spin-related observables in processes involving polarized
protons provides to probe correlation between the proton's spin and the spins
of its constituents.  There have been several investigations of polarized
parton distributions inspired by the EMC measurement [1] of the polarized
structure function $g_1^p$.  Together with the latest available data, the
recent SMC measurement [2] of the integrated proton structure
function $g_1^p$ is
\begin{equation}
\int_0^1 dx\, g_1^p (x)~(SMC/EMC/SLAC)~~=~~0.142\pm 0.008\pm 0.011 ~ (Q^2=10
GeV^2).
\end{equation}

In the na{\"\i}ve parton model, $g_1^p$ is related to the polarized quark
densities ($q_i^\uparrow,q_i^\downarrow$) with spin parallel or antiparallel
to the longitudinally polarized parent proton (at momentum transfer scale
$Q^2$)
\begin{equation}
 g_1^p (x,Q^2)~~=~~{1\over2}\sum_qe_q^2 \left[\Delta q (x,Q^2) +\Delta \bar{q}
 (x,Q^2)\right],
\end{equation}
where $e_q$'s are the quark charges and $\Delta q (x,Q^2) = q^\uparrow
(x,Q^2) - q^\downarrow (x,Q^2)$.  When combined with the values of
the parameters $F$ and $D$  determined from low-energy neutron and
hyperon beta decays by means of SU(3) flavor symmetry, the first moment
of the flavor singlet part of $g_1^p$ is very close to zero.  This
implies that the total fraction of the proton helicity carried by
all quarks and antiquarks almost vanishes, in contradiction to previous
na{\"\i}ve expectation [3].
Suggestions have been made to explain this startling conclusion
in terms of a large negative polarization of the sea quarks inside the
proton [4] or a large positive polarization for the gluons [5] or a
suitable combination of both [6].

However, the result of recent NMC measurement [7] hints that the $u$-$d$
flavor symmetry of the sea is broken.  This hint comes from the
violation of the Gottfried sum rule.  The NMC obtained
the Gottfried sum to be $S_G = 0.235 \pm 0.026 \, (Q^2 = 4 GeV^2)$.
Under the assumptions of isospin symmetry for the nucleon and for the sea quark
distributions in the proton, the Gottfried sum reads
\begin{eqnarray}
 S_G &=& \int_0^1{dx\over x}\left[F_2^{\mu p}(x)
                            -F_2^{\mu n}(x)\right]\nonumber\\
     &=& {1\over 3}\int_0^1dx\left[u(x)+{\bar u}(x)-d(x)
                            -{\bar d}(x)\right]\nonumber\\
     &=& {1\over 3}\,.
\end{eqnarray}
This implies a large SU(2) flavor symmetry breaking in the quark sea or the
suppression for the production of $u{\bar u}$ pairs relative to $d{\bar d}$
pairs in the proton:
\begin{eqnarray}
 d_s - u_s & = & {\bar d}-{\bar u} \nonumber\\
           & = & \int_0^1 dx\left[{\bar d}(x)-{\bar u}(x)\right] \nonumber\\
           & = & 0.147\,.
\end{eqnarray}
In this paper, we aim at understanding the spin content of the polarized
sea quarks and providing an ingredient in decoding the spin puzzle.

Let us consider the first moment of the spin-dependent structure function
in the parton model with QCD corrections.  It is given entirely by the
proton matrix elements $a_0, a_3$ and $a_8$ of the axial
vector currents multiplied by the relevant Wilson coeffients [8]
\begin{eqnarray}
\int_0^1 dx\,g_1^p (x,Q^2) &=& {1\over12}(1-{\alpha_s(Q^2)\over\pi})a_3
+{1\over36} (1-{\alpha_s(Q^2)\over\pi})a_8
\nonumber\\
& & + {1\over9}\left(1-{\alpha_s(Q^2)\over\pi}({33-8N_f\over33-2N_f})\right)a_0
\nonumber\\
& & - {\alpha_s(Q^2)\over6\pi} \Delta G(x,Q^2)\,,
\end{eqnarray}
where the gluon contributes to $g_1^p$ via the $\gamma_5$-triangle
anomaly [5, 9].  The matrix elements can be decomposed into non-singlet and
singlet polarized quark distributions.  In terms of quark polarizations
$\Delta q \equiv \int dx(q^\uparrow (x) - q^\downarrow (x))$, or
the $F$-type and $D$-type couplings, the matrix elements read
\begin{eqnarray}
a_3 &=& \Delta u + \Delta\bar{u} - \Delta d - \Delta\bar{d}~~\equiv~~ F + D
\,,\nonumber\\
a_8 &=& \Delta u + \Delta \bar{u} + \Delta d + \Delta\bar{d} - 2\Delta s
- 2\Delta\bar{s}~~ \equiv~~ 3F - D\,,\\
a_0 &=& \Delta u + \Delta \bar{u} + \Delta d + \Delta\bar{d} + \Delta s
+ \Delta\bar{s}~~\equiv~~2S_z^{quarks}\,, \nonumber
\end{eqnarray}
where the singlet part $a_0$ corresponds the total spin carried by
all the quarks.

We assume that there is no significant intrinsic
polarization of the gluon ($\Delta G = 0$).  Equating (5) to (1) and taking
$N_f = 3$ and $\Lambda_{QCD} = 177 MeV$ for (5)
at the starting point of our perturbative evolution $Q^2 = 4 GeV^2$, we
obtain
\begin{eqnarray}
 \Delta u_v &=& 2F + 2\Delta s - 2\Delta u_s~~=~~0.825 - 2\Delta u_s
\,,\nonumber\\
 \Delta d_v &=& F - D + 2\Delta s - 2\Delta d_s~~=~~-0.432 - 2\Delta d_s\,,\\
 \Delta s &=& -0.047\pm0.025\,,\nonumber
\end{eqnarray}
where $\Delta \bar{u}=\Delta u_s$, $\Delta \bar{d}=\Delta d_s$ and $\Delta
\bar{s}=\Delta s$ are assumed and might be considered since there is also
a contribution to the proton's spin from the orbital angular momentum
of the partons.  In (7) we have used $F=0.459\pm0.008$ and $D=0.798\mp0.008$
from ref.[10].  The general spin decomposition of the proton is
( in the Skyrme model [4, 11] and its simple
extensions, there are no gluons ($\Delta G = 0$))
\begin{equation}
 {1\over2}~~=~~S_z^{quarks} + \Delta G(= 0) + <L_z> .
\end{equation}
The compensation between the quark helicities and the orbital angular momentum
to the proton in the infinite momentum frame is consistent with QCD in
such a way that the proton's spin is unchanged.  We assume that the
polarized sea distributions for $u, d$ and $s$ quarks are different.
To a good approximation, the Pauli exclusion principle guides us in making
reasonable assumptions for various polarized parton distributions.  Using
(7) to relate the valence distributions to the values of the axial
coupling at $Q^2 = 4 GeV^2$, we have
\begin{eqnarray}
 u_v^\uparrow &=& 1 + F + \delta\,\,, ~~ u_v^\downarrow~~=~~1 - F - \delta\,,
\nonumber\\
 d_v^\uparrow &=& {1 + F - D\over2} + \beta\,\,, ~~ d_v^\downarrow~~=~~
{1 - F + D\over2} - \beta\,,
\end{eqnarray}
where
\begin{eqnarray}
 \delta &=& \Delta s - \Delta u_s\, ,\nonumber\\
 \beta  &=& \Delta s - \Delta d_s\, .
\end{eqnarray}
When $Q^2 \to 0\,GeV^2$, no contribution from the sea $(\delta = \beta =
0)$, one has $\Delta u_v = 2F$ and $\Delta d_v = F - D$ [12].  The Pauli
principle might be advocated to account for the distribution of the $u$
quark with respect to that of the $d$ quark in the proton.  Therefore, we
rewrite $u$ and $d$ in (9) as
\begin{eqnarray}
 u_v^\downarrow &=& ( 1 - F - \delta ) d_v\, ,\nonumber\\
 d_v^\downarrow &=& ( {1 - F + D\over2} - \beta ) {u_v\over2}\,,
\end{eqnarray}
and generalize the above relations to the sea quark distributions
\begin{eqnarray}
 u_s^\downarrow &=& ( 1 - F -\delta ) d_s\, ,\nonumber\\
 d_s^\downarrow &=& ( {1 - F +D\over2} - \beta ) {u_s\over2}\,.
\end{eqnarray}
We first start from $\Delta G = 0$ and consider an ansatz for
spin-dependent quark distributions based on SU(3) flavor symmetry
breaking effect for polarized sea.
{}From (4), (10) and (12), we obtain
\begin{eqnarray}
 \Delta u_s &=& \Delta \bar{u}\nonumber\\
            &=& u_s - 2u_s^\downarrow\nonumber\\
            &=& {-0.147 + (-1 + 2F + 2\Delta s)d_s\over1 + 2d_s}\,,
\end{eqnarray}
and, similarly,
\begin{eqnarray}
 \Delta d_s &=& \Delta \bar{d}\nonumber\\
            &=& {0.294 + (1 + F -D +2\Delta s)u_s\over2 + 2u_s}\,,
\end{eqnarray}
where $\bar{u}^\uparrow = u_s^\uparrow$, $\bar{u}^\downarrow = u_s^
\downarrow$, $\bar{d}^\uparrow = d_s^\uparrow$ and $\bar{d}^\downarrow
= d_s^\downarrow$ are assumed.

In the following analysis we shall use the set MRS(A) of unpolarized parton
distribution functions given in ref. [13].  This set is
extracted at the reference scale $Q^2 = 4 GeV^2$ from several
experiments.  The values of $d_s$ and $u_s$ can be obtained by the
integrated MRS(A) set with the distributions of the $c$ quarks in the
proton being set identically to zero.  From (7), (13) and (14), the
normalizations
of the polarized sea-quark distributions are given by
\begin{eqnarray}
\Delta u_s &=& -0.093\,,~~ \Delta d_s ~~=~~ 0.262\,,\nonumber\\
\Delta u_v &=& 1.011\,,~~~~\, \Delta d_v ~~=~~ -0.956\,.
\end{eqnarray}
Including QCD corrections, we get $\Delta d_s
\approx -2.8 \Delta u_s \approx -5.6 \Delta s $ and $S_z^{quarks}= 0.149
\pm0.037 $ and most of the proton's spin come from the orbital angular
momentum $L_z$.  A nonzero $L_z$ has the phenomenological consequence that
there is an intrinsic transverse momentum, since in the na{\"\i}ve parton
model with all partonic momentum parrallel to the parent's momentum, one
has $L_z=0$.  Contrary to SU(3) invariance of sea polarization, (15)
implies that $d_s$ quarks prefer to polarize in the same direction of
the proton's spin, while $u_s$ and $s$ quarks do not.  Perturbative QCD
arguments [14] suggest that the valence quarks at $x = 1$ remember the
spin of the parent proton, and the quark distributions are dominated by
the sea as $x \rightarrow 0$.  We shall take the following form for polarized
valence-quark distributions [15]:
\begin{equation}
\Delta q_v(x)~~=~~\left({x-x_0\over1-x_0}\right)x^p q_v(x)\,,
\end{equation}
which satisfies the boundary conditions and $x_0$ is a free
parameter.  The sign of $\Delta q_v(x)$ is flipped at $x = x_0$ when $p$
is positive.  Since $\Delta u_v$ is positive, we take $x_0 = 0$ and write
the parametrization for $\Delta u_v$ as $\Delta u_v(x) = x^{\gamma_{u_v}}
u_v(x) $.  While for $\Delta d_v$, we have $\Delta d_v < 0$, the helicity
of a $d_v$ quark relative to its parent proton is flipped.  Using the
normalizations (15), we obtain the
parametrizations
\begin{eqnarray}
\Delta u_v(x) &=& x^{0.275}u_v(x)\,,\nonumber\\
\Delta d_v(x) &=& \left({x-0.9\over1-0.9}\right)x^{0.818} d _v(x)\,.
\end{eqnarray}
In order to ensure that the positivity constraints
($\vert\Delta q_s(x)\vert \le q_s(x)$)
be satisfied for $0\leq x \leq 1$, we use the simple parametrizations
$\Delta q_s(x) = {\pm} x^{\gamma_{q_s}} q_s(x)$,  where the ``+'' sign is for
$\Delta q_s > 0$ and the ``--'' sign is for $\Delta q_s < 0$.
The $\gamma_{q_s}$'s are given by
\begin{eqnarray}
\gamma_{u_s} &=& 0.693\,, ~~ \gamma_{d_s}~~=~~0.521\,,\nonumber\\
\gamma_s &=& 0.722\,.
\end{eqnarray}
One observes that one can use the general parametrization form
$\Delta q_s(x)\approx x^m(1-x)^nq_s(x)$, but this does not make
any practically different distributions since the behavior of
sea quarks is considered to be dominated at small $x$.

All quark spin distributions are determined initially at
$Q^2 = 4 GeV^2$ and are used as the input distributions, from which
we obtain the distributions at higher $Q^2$ values by a numerical
integration [16] of the Altarelli-Parisi (AP) equations for the polarized
case [17].  With three different $Q^2$ values, the theoretical fits to
the data on $xg_1^p(x)$ from the EMC and SMC [1, 2] are shown in
Fig. 1.  Fig. 1 shows that the $Q^2$ variation has small effect on
$xg_1^p(x)$.  Fig. 2 exhibits the $Q^2$ evolution of the polarized
sea-quark densities.  The
distributions $\Delta u_s(x)$ and $\Delta s(x)$
increase while $\Delta d_s(x)$ decreases with $Q^2$.  In Fig. 2,
$\Delta d_s (x)$ drastically differs from $\Delta u_s (x)$ and
$\Delta s (x)$.  We may regard the polarized quark distributions as having
the Pauli exclusion principle reflected by the SU(3) flavor symmetry
breaking for the sea polarization.  Likewise, shown in Fig. 3 is
the $x$ behavior of the spin-dependent gluon densities, which are
dynamically generated from the bremsstrahlung of the quarks.  These
different-$Q^2$ behaviors indicate that if the gluon is initially
unpolarized at the perturbative evolution scale then the induced gluon
polarization will be generated at larger $Q^2$.  The gluon spin does
increase with $lnQ^2$ in Fig. 4, but the quark-induced $\Delta G$ is
small.  The resulting $\Delta G$ is sensitive to the initial shape of
the gluon distribution, which gives a contribution to $g_1^p$.  Fig. 5
shows that the net quark polarization is nearly independent of $Q^2$.

In this paper we have obtained a consistent set of spin-dependent quark
distributions which provide a tool for deep inelastic structure function
measurements and theoretical calculations of cross sections in polarized
hadron-hadron and lepton-hadron collisions.  One may wish to consider
a non-zero polarized charm sea for $Q^2 > m_c^2$ and to make a change of
basis from $a_3$, $a_8$ and $a_0$ to $a_3$, $a_8$, $a_{15}$($= \Delta u
+\Delta \bar u +\Delta d +\Delta \bar d +\Delta s +\Delta \bar s - 3\Delta c
- 3\Delta \bar c$) and $a_0$($= \Delta u +\Delta \bar u +\Delta d +\Delta
\bar d + \Delta s +\Delta \bar s +\Delta c +\Delta \bar c$)
at the initial $Q^2$.  In practice, the non-perturbative
contribution of the polarized c-quark densities is too small to distort
the analysis at large $x$, but it does modify the normalization
and the shape of the polarized parton distributions in the small $x$ region.
It is also interesting to note that there are $\bar{s}s$ pairs
in the proton due to $\Delta s \neq 0$.  One can extend the analysis to
dynamical processes which involve OZI-forbidden diagrams.
Phenomenologically, the spin-dependent parton densities improve
the description of the polarized nuclear structure functions but we need
more accurate experimental data on the nucleon spin to
understand the spin-transfer mechanism and to clarify how the proton's
spin is distributed among its valence quarks and the flavored $q\bar q$ sea.

\vskip 1cm
\noindent{\large \bf Acknowledgments}

\noindent
I should like to thank Prof. Su-Long Nyeo for useful discussions.
This research was supported by the National Science Council
of the Republic of China under Contract No.~NSC 83-0208-M-006-057.

\newpage
\noindent{\large \bf Figure Captions}

\noindent
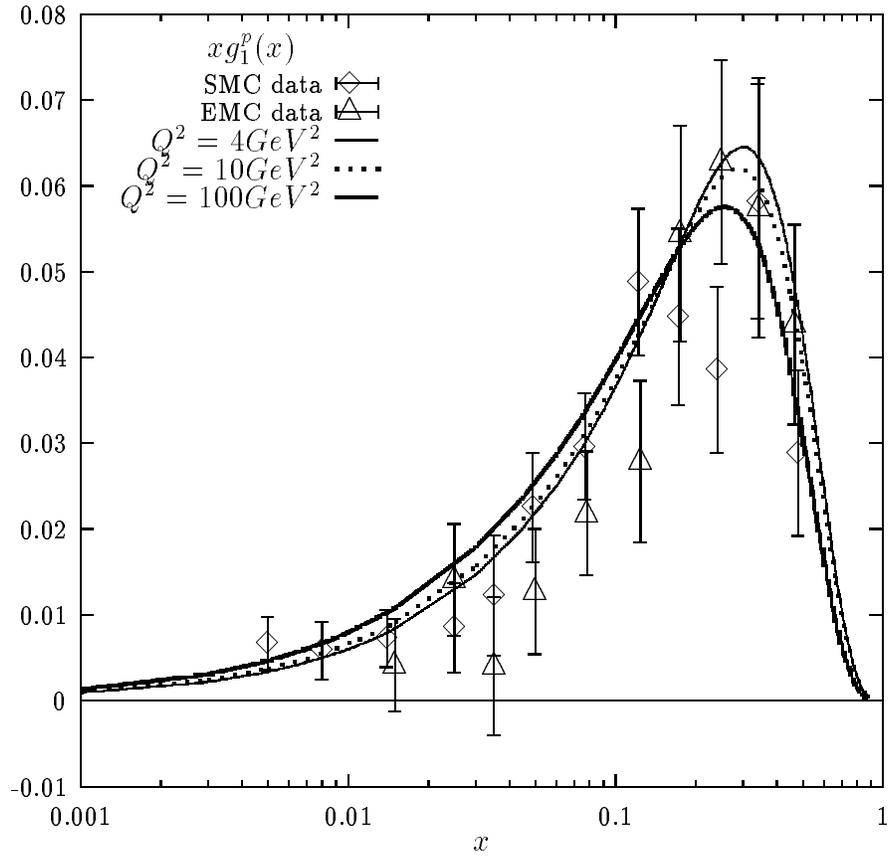
Figure 1. Fits to the $g_1^p(x)$ structure function at three values of $Q^2$.

\noindent
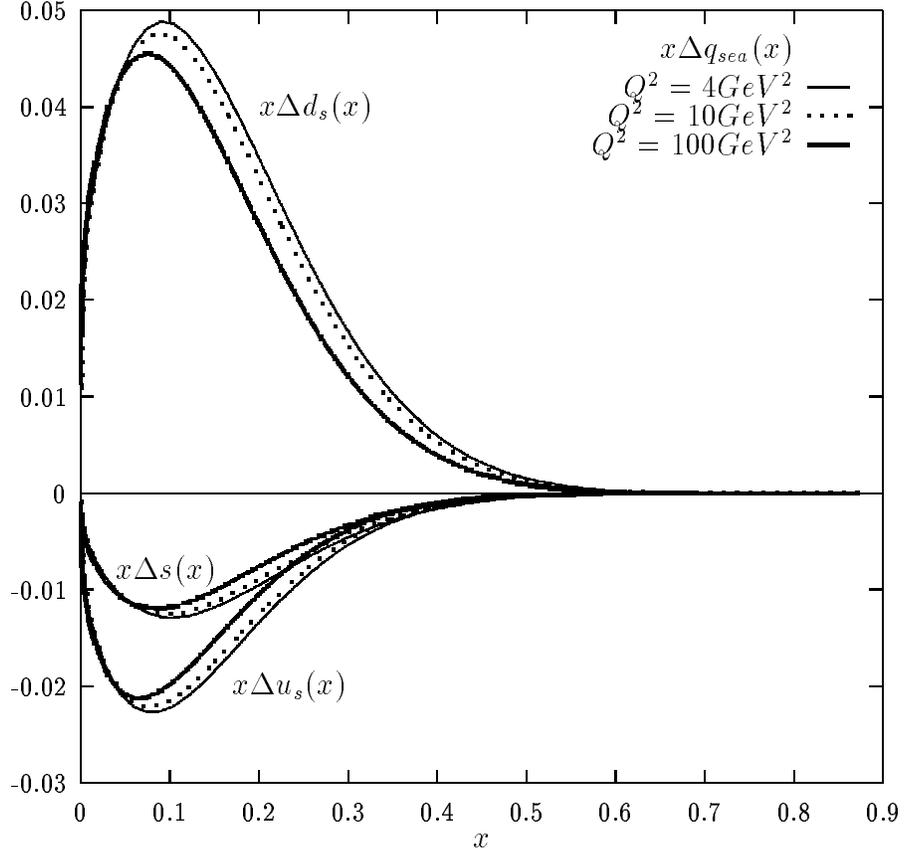
Figure 2. The evolution of polarized sea densities at three values of $Q^2$.

\noindent
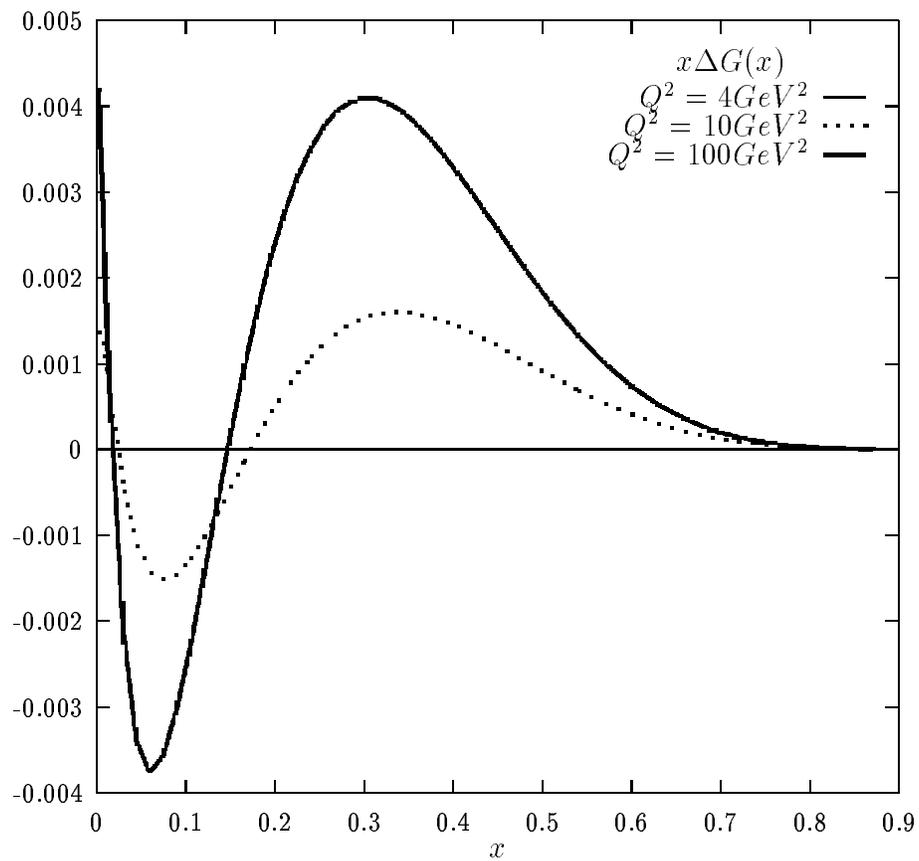
Figure 3. The evolution of polarized gluon densities at three values of $Q^2$.

\noindent
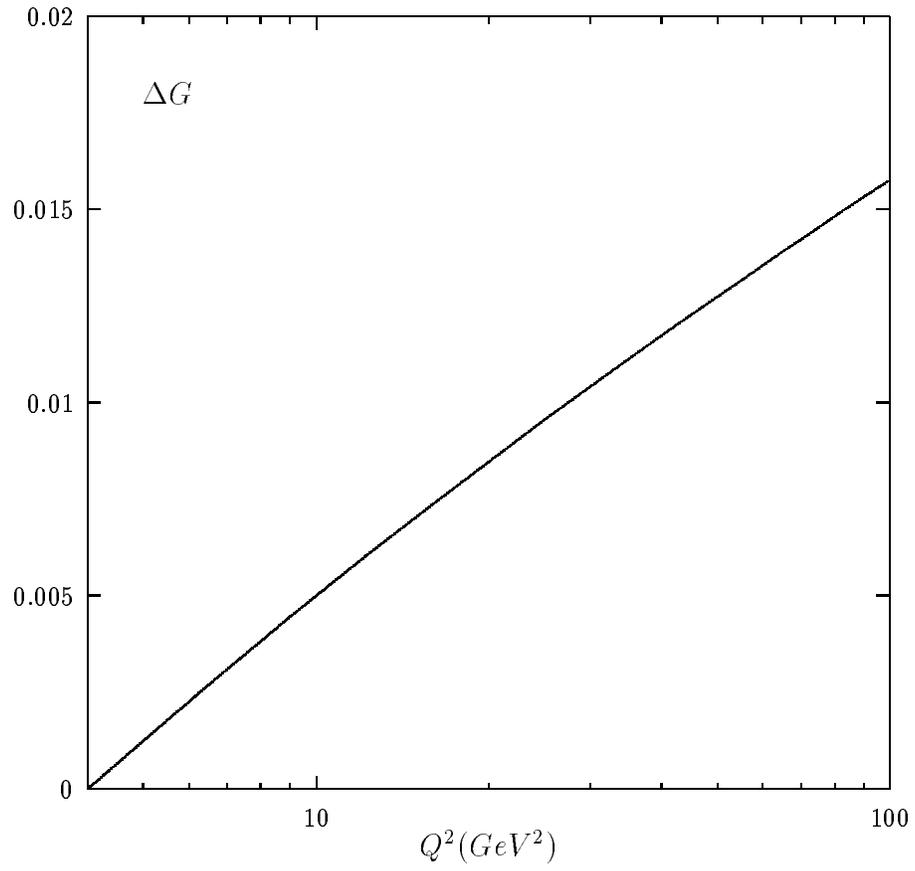
Figure 4. The $x$-integrated distribution $\Delta G$ over the range
$10^{-4} < x < 1$.

\noindent
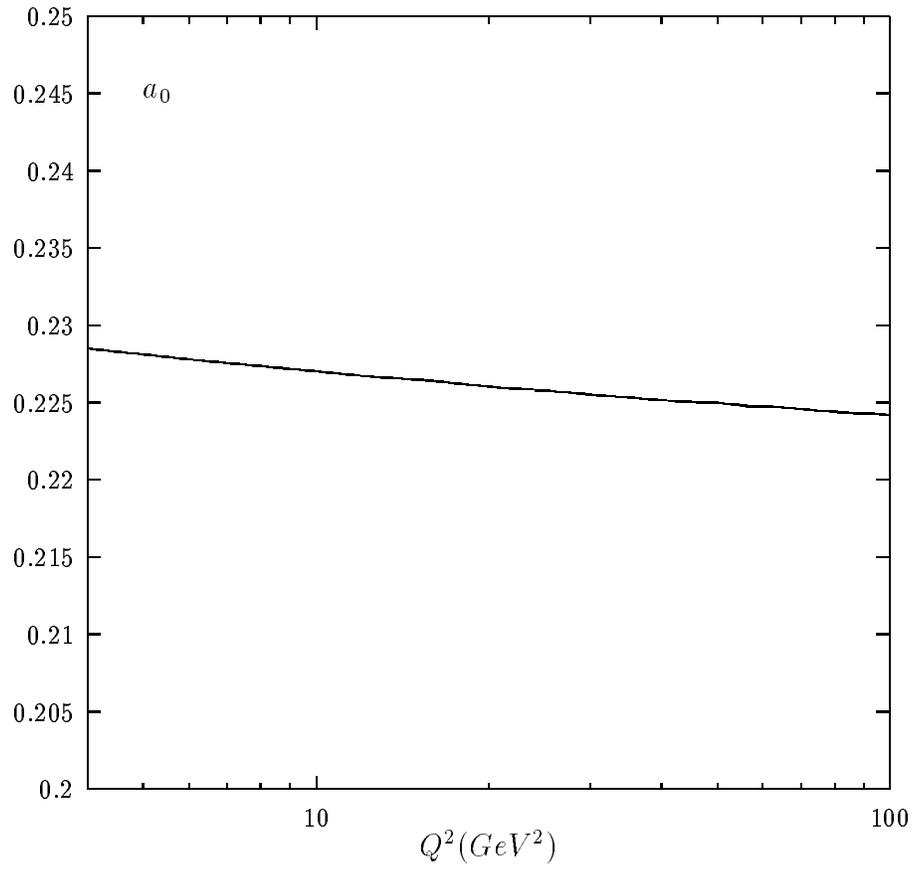
Figure 5. The $x$-integrated distribution $a_0$ over the range
$10^{-4} < x < 1$.

\end{document}